# DCAS: Decoupling CLI Agent Scaffolding to Internalize Planning across Scaffolds

Kishanthan Thangarajah
Centre for Software Excellence,
Huawei Canada
Markham, Canada
kishanthan.thangarajah1@huawei.com

Boyuan Chen
Centre for Software Excellence,
Huawei Canada
Markham, Canada
boyuan.chen1@huawei.com

Ahmed E. Hassan
Queen's University
Kingston, Canada
ahmed@cs.queensu.ca

## Abstract

CLI-based software-engineering agents have matured rapidly, yet the open ecosystem has converged on a single training environment: trajectory datasets used to fine-tune open models are collected almost exclusively under OpenHands. Models fine-tuned on this data score well under OpenHands but degrade substantially when deployed under any non-training scaffold. Untrained base models do not show this divergence, indicating the gap is fine-tuning-induced and tied to the conventions of the training scaffold. We argue that a load-bearing scaffold-specific behavior is *planning structure*, in two senses this paper distinguishes: *explicit planning*, a pre-execution plan produced as a first-class artifact, and *implicit planning*, the structural conventions that shape execution throughout the agent loop. Under this hypothesis, closing the gap requires moving planning from a fixed scaffold artifact to a learned model capability. We introduce *Decoupling CLI Agent Scaffolding (DCAS)*, a backend-substitution interception layer that routes API traffic between any CLI scaffold and any backend model without modifying the scaffold, enabling cross-scaffold evaluation and planning-aware trajectory collection. Using DCAS, a controlled plan-source intervention confirms planning quality is a high-leverage component, with gains exceeding the cross-scaffold drops we observe. A model fine-tuned on a small set of DCAS-collected planning-aware trajectories under a single scaffold gains consistently across non-training scaffolds, and the two senses of planning are empirically separable in training data.

## CCS Concepts

• **Software and its engineering → Software development techniques**; • **Computing methodologies → Machine learning**.

## Keywords

CLI agents, software engineering agents, strategic planning, supervised fine-tuning, LLM agents, scaffold decoupling

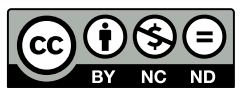

## 1 Introduction

CLI-based software-engineering agents have advanced significantly, with open models at the 32B scale now resolving more than half of SWE-bench Verified instances [2, 16, 21, 24]. Yet the open ecosystem has converged on a single training environment: SWE-Gym [19], Nebius [28], SWE-Lego [24], and CoderForge [2] all collect their training trajectories exclusively under OpenHands [31]. Models fine-tuned on this data score well under OpenHands but degrade substantially when deployed under any non-training scaffold (Section 2). Practitioners, however, select CLI scaffolds based on cost, licensing, latency, and data-privacy constraints [7, 8], not on which scaffold the available open models happen to have been trained under. The cause of this gap, and whether it can be closed without scaffold-specific retraining, has not been studied.

The divergence is not a property of the underlying model. Untrained base models vary far less across evaluated scaffolds than the models fine-tuned from them; only after single-scaffold fine-tuning does the spread widen sharply. Whatever causes the drop is therefore something fine-tuning installs. We hypothesize it is tied to the conventions of the training scaffold rather than to model capability. This narrows the search: the responsible behavior must be one that scaffolds visibly differ in *and* that fine-tuning teaches a model to follow.

Recent evidence from Liu et al. [13] narrows the search further: across 16,991 CLI agent trajectories, plan quality and plan adherence are the primary drivers of resolution rate. We therefore focus on planning, and use the term in two senses this paper distinguishes. *Explicit planning* is a pre-execution step in which the model produces a plan as a first-class artifact before acting; some scaffolds surface this as a discrete planning phase, exemplified by Claude Code's Plan Mode [1] and OpenCode's Plan agent [22]. *Implicit planning* is the structural behavior every scaffold imposes on the agent loop turn by turn: how work is decomposed into sub-steps, when exploration gives way to action, how tool calls are sequenced, how failures trigger replanning. Scaffolds therefore differ on both axes: some expose an explicit planning phase and others do not, and we expect the implicit planning each scaffold imposes to take its own characteristic form. Our hypothesis, then, is that fine-tuning under one scaffold installs that scaffold's particular blend of the two, and that deployment under another exposes the mismatch.

Underneath their surface differences, however, CLI scaffolds share a similar ReAct-style [38] act/observe execution loop. A model trained under one scaffold therefore has the *capacity* to operate under others; we argue that what it lacks under a non-training scaffold is familiarity with that scaffold's planning conventions. If the hypothesis is right, it follows that a model can be brought to scaffold-portable performance by training on a small amount of trajectories that capture both senses of planning under scaffolds the model is not yet familiar with.

Testing this hypothesis requires three capabilities that existing tooling does not provide. First, the same backend model must be routable through any target scaffold with that scaffold's full convention surface intact, for controlled cross-scaffold evaluation. Second, trajectories must be collectable in a form that captures both explicit and implicit planning under any chosen scaffold, including scaffolds whose proprietary backends are not normally substitutable. Third, fine-tuning must operate on those trajectories without modifying the scaffold itself. Each requirement is currently blocked. Public trajectory datasets bundle planning and execution opaquely under a single scaffold; scaffolds with rich planning behavior are tightly coupled to proprietary backends; cross-scaffold evaluation across incompatible API formats has no general-purpose infrastructure.

We introduce *Decoupling CLI Agent Scaffolding (DCAS)*, a backend-substitution interception layer that routes API traffic between any CLI scaffold and any backend model without modifying the scaffold. DCAS enables cross-scaffold evaluation of the same backend model and planning-aware trajectory collection under any target scaffold. Using DCAS, we structure the test of the hypothesis across three research questions:

- **RQ1.** Does explicit planning quality move task performance? We hold the executor model and scaffold constant and vary only the source of the plan, isolating planning's contribution under controlled conditions.
- **RQ2.** Can planning be internalized through fine-tuning? We compare plan-only and plan-and-execution training to test which sense each composition installs.
- **RQ3.** Does the learned planning capability generalize to scaffolds the model never saw during training? A structural skill should transfer; scaffold-specific memorization should not.

In addressing these questions, this paper makes the following contributions:

- **DCAS**, the backend-substitution interception layer that enables the rest of this paper's experiments (Section 3).
- **Effect of explicit planning quality (RQ1).** Holding the executor model and scaffold fixed, varying only the plan source produces gains on SWE-bench Verified whose magnitude exceeds the cross-scaffold drops we observe, confirming planning is a high-leverage component (Section 4).
- **Planning as a learnable, transferable capability (RQ2 and RQ3).** A model fine-tuned on a small set of DCAS-collected planning-aware trajectories under a single scaffold gains consistently across multiple non-training scaffolds (Sections 5 and 6).

The remainder of this paper is organized as follows. Section 2 presents the background and motivating study that establishes the cross-scaffold gap, narrows its cause, and develops the explicit and implicit planning distinction with empirical support. Section 3 describes the DCAS design and experimental setup. Sections 4, 5, and 6 report RQ1, RQ2, and RQ3. Section 7 discusses implications, Section 8 reviews related work, Section 9 discusses threats to validity, and Section 10 concludes.

## 2 Background and Motivation

This section establishes the empirical phenomenon the paper investigates and develops the hypothesis that the rest of the paper tests.

### 2.1 CLI Agent Scaffolds and the Training Landscape

A CLI agent scaffold is the harness that turns a language model into an autonomous coding agent: it manages the agent loop, exposes a set of tools to the model, structures multi-turn conversations, and decides when the agent has finished. Four scaffolds appear in this study, chosen to span the spectrum of design choices currently in deployment.

**OpenHands** [31] is the dominant open scaffold and the primary collection environment for the open trajectory ecosystem. SWE-Gym [19], Nebius [28], SWE-Lego [24], and CoderForge [2] all collect under OpenHands. The practical consequence is that almost every publicly available open model with strong CLI agent performance is predominantly trained on OpenHands trajectories.

**Claude Code** [1] and **OpenCode** [22] are frontier-style terminal scaffolds that incorporate a dedicated planning phase: the agent is expected to produce a structured plan before any code modification, and tool schemas, turn management, and context handling are designed around this two-phase decomposition. Claude Code is closed-source and tightly coupled to Anthropic models; OpenCode is open-source and provider-agnostic.

**mini-swe-agent** [12] is a deliberately minimal ∼100-line scaffold with bash-only interaction, no dedicated planning phase, no structured turn management, and no tool-calling interface. It serves as a stress test: a scaffold that is structurally unlike Claude Code in almost every way, and on which any cross-scaffold generalization claim must hold up.

These scaffolds differ most visibly in how they handle planning. Claude Code and OpenCode embed a discrete planning phase in their agent loop. OpenHands embeds planning differently, distributed across its CodeAct cycle [30] rather than concentrated in a dedicated stage. mini-swe-agent has no explicit planning structure at all. Sub-step granularity, tool-chaining conventions, and failure-recovery patterns vary correspondingly. We return to these structural differences in Section 2.3.

### 2.2 The Cross-Scaffold Deployment Gap

OpenHands is the predominant collection environment for open trajectory datasets (Section 2.1), so models fine-tuned on this data may carry conventions that do not transfer. We test this on four top-performing 30B–32B open models evaluated on SWE-bench Verified [6] across all four scaffolds, alongside the untrained base models they were fine-tuned from. Two are trained under OpenHands and two under other pipelines, so the question extends beyond OpenHands-specific training. Table 1 reports the results.

**Table 1: Pass@1 on SWE-bench Verified, max 100 turns. Italicized rows are untrained base models. † reported result; all others are our runs. OH = OpenHands (v0.50.0), CC = Claude Code, OC = OpenCode, Mini-SWE = mini-swe-agent.**

| Model | OH | CC | OC | Mini-SWE |
|---|---|---|---|---|
| *Qwen3-32B* (base) | 29.0 | 23.2 | 18.4 | 8.0 |
| SWE-Lego-Qwen3-32B [24] | 52.6† | 44.2 | 8.4 | 37.6 |
| SERA-32B [21] | 54.2†[a] | 33.4 | 35.8 | 26.4 |
| Qwen3-32B-Nex-N1 [16][b] | 50.5† | 50.8 | 26.2 | 15.6 |
| *Qwen3-30B-A3B-Instruct-2507* (base) | 25.2† | 26.8 | 27.6 | 14.0 |
| Nebius-SWE-Rebench-30B [28] | 49.7† | 46.0 | 35.8 | 20.4 |

[a]Public score at 64K context; model trained to 32,768 tokens.
[b]Trained on diverse multi-format trajectories [16], not a single scaffold.

**Every fine-tuned model degrades on at least one scaffold it was not trained on.** The degradations fall into three categories. *Score degradation with scaffold distance* is the most common: SWE-Lego-Qwen3-32B drops from 52.6% on OpenHands to 44.2% on Claude Code, and Nebius-SWE-Rebench-30B drops from 49.7% on OpenHands to 20.4% on mini-swe-agent, a 29-point collapse. *Tool-call format incompatibility* produces sharper failures: SWE-Lego-Qwen3-32B reaches only 8.4% on OpenCode because its tool-call output format is incompatible with OpenCode's parser, leaving most trajectories with zero parseable actions. *Scaffold-specific context constraints* produced complete failures before workarounds were applied: both SERA-32B and Qwen3-32B-Nex-N1 are fine-tuned to a context length that conflicts with Claude Code's hard-coded ceiling, causing requests to fail under Claude Code but not OpenHands. We resolved this through context-length scaling and report the post-fix scores.

**The divergence is fine-tuning-induced.** The italicized rows in Table 1 report the same evaluation on the untrained base models. Both Qwen3-32B (29.0/23.2/18.4/8.0) and Qwen3-30B-A3B-Instruct-2507 (25.2/26.8/27.6/14.0) score comparably across the four scaffolds within their respective families. The spread within each base-model row is smaller than the spread within any fine-tuned row. Whatever causes the post-fine-tuning divergence is therefore something installed by training, not a property of the underlying model. The mechanism must be a behavior that scaffolds visibly differ in and that fine-tuning teaches a model to follow.

**Broader training distribution does not close the gap.** Qwen3-32B-Nex-N1 was trained on diverse multi-format trajectories [16] rather than under a single scaffold. It narrows the gap on Claude Code (50.8%, comparable to its OpenHands score) but still degrades substantially on OpenCode (26.2%) and mini-swe-agent (15.6%). Simply broadening the training distribution by mixing scaffolds is therefore not sufficient; the dependence on training-scaffold conventions persists.

## 2.3 Planning Structure as the Candidate Mechanism

The diagnostic evidence in Section 2.2 narrows the search to scaffold-specific behaviors that fine-tuning installs. We argue that planning structure is a load-bearing one and the one with the greatest leverage, in the two senses introduced in Section 1: *explicit planning*, the pre-execution step in which the model produces a plan as a first-class artifact before acting, and *implicit planning*, the structural behavior every scaffold imposes on the agent loop turn by turn (sub-step granularity, when exploration gives way to action, tool-call sequencing, failure recovery). The cross-scaffold gap operates through both, and the rest of the paper teases them apart empirically.

The argument for planning rests on one inferential move. Scaffolds differ most visibly in their planning structure (Section 2.1), but underneath they share a similar execution loop: all four scaffolds in our study run a ReAct-style [38] act/observe loop in which the model emits an action or a planning step, the scaffold returns an observation, and the loop continues until completion. A model trained under one scaffold therefore has the *capacity* to operate under any of them; what it lacks under a non-training scaffold is familiarity with that scaffold's planning conventions. If the underlying loop is shared, the cross-scaffold gap cannot be a capability deficit, and planning conventions become the most plausible locus of the mismatch. Liu et al. [13] provide independent corroboration: across 16,991 CLI agent trajectories, plan quality and plan adherence are the primary drivers of resolution rate.

Before designing the controlled experiments in Sections 4 to 6, we ran a simple check on the very models in Table 1: does activating even an explicit planning step improve cross-scaffold scores? Table 2 shows that a self-generated planning turn produces consistent gains on both Claude Code and OpenCode for two representative models, despite no targeted training. The gains are modest, but they show planning structure is at minimum a movable variable on these scaffolds, motivating the controlled intervention in RQ1 (Section 4).

**Table 2: Planning-stage activation on representative fine-tuned models. Pass@1 on SWE-bench Verified. Teal deltas are gains over the no-plan score in Table 1.**

| Model | Planning | CC (%) | OC (%) |
|---|---|---|---|
| Nebius-SWE-Rebench-30B | No plan | 46.0 | 35.8 |
| | Self-plan | 48.4 (+2.4) | 38.0 (+2.2) |
| Qwen3-32B-Nex-N1 | No plan | 50.8 | 26.2 |
| | Self-plan | 52.8 (+2.0) | 28.0 (+1.8) |

The cross-scaffold deployment gap is, to a substantial degree, a planning-convention mismatch: a model trained under one scaffold has internalized one combination of explicit and implicit planning conventions, and deployment under another exposes that combination as scaffold-specific. The implication, if the hypothesis holds, is that planning needs to move from a fixed scaffold artifact to a learned model capability. Sections 4 to 6 test the hypothesis.

# 3 DCAS: Design and Experimental Setup

This section describes the DCAS interception layer and the experimental setup used to test the planning hypothesis stated in Section 2.3.

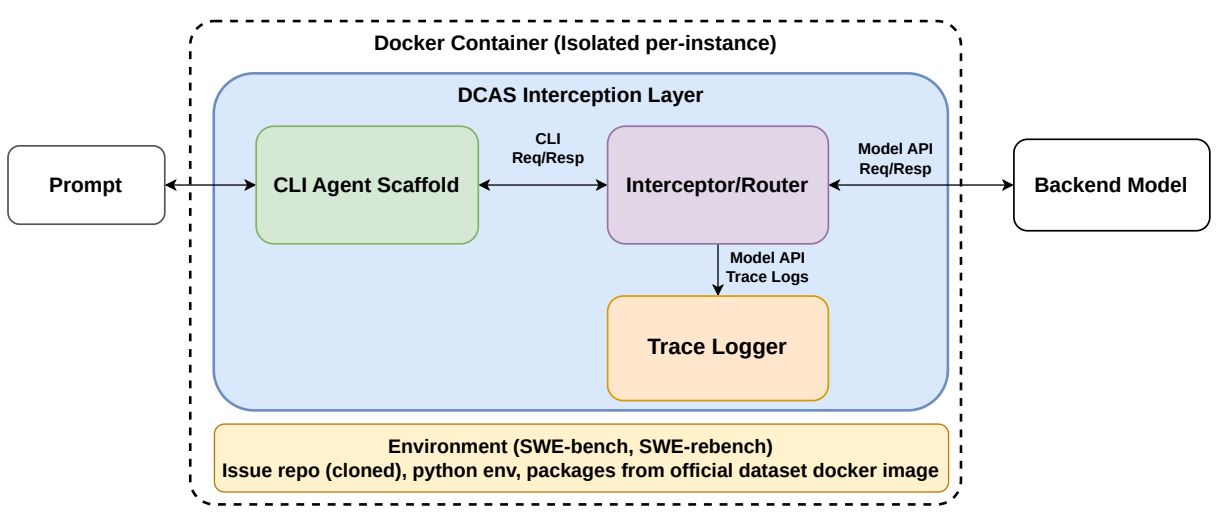

**Figure 1: The DCAS interception layer routes API traffic between any CLI scaffold and any substituted backend model without modifying the scaffold. The trace logger records all interactions as trajectory files for SFT data collection.**

## 3.1 The DCAS Interception Layer

DCAS operates by routing API traffic between a CLI agent scaffold and its backend model through an intermediary that translates requests and responses to match the target model's format, allowing any backend model to be substituted while the scaffold remains entirely unmodified. The DCAS interception layer was inspired by and built upon the open-source SWE-bench runner by McMillan [15], which demonstrated the feasibility of intercepting CLI scaffold API traffic for multi-model evaluation, and which we extended with planning-stage separation, structured trajectory logging, and multi-scaffold support.

In this work we apply DCAS to Claude Code as the primary training and evaluation scaffold. All experiments use CC 2.0.76, the stable Claude Code release available at the start of the study, kept fixed throughout for consistency. For RQ3 version generalization, we additionally use CC 2.1.73, a newer release, to test whether the learned planning capability transfers across scaffold versions. The open scaffolds evaluated in RQ3 are OpenCode v0.0.55 [22] and mini-swe-agent v2.2.8 [12]; the same versions appear in Table 1 (Section 2.2) and throughout this paper. OpenCode is a widely adopted open-source terminal CLI agent with multi-provider model support. mini-swe-agent is a minimal bash-only scaffold with no dedicated planning phase, no structured turn management, and no tool-calling interface, making it a useful stress test for scaffold generalization.

Because the scaffold is identical across all experimental configurations within each RQ, any observed performance difference between conditions is attributable to the model or the plan, not to scaffold variation.

## 3.2 The Planner-Executor Setup

For the controlled planning experiments in RQ1, plans are provided to the backend model through the user prompt. Three prompt templates govern the three experimental conditions.

In the **no-plan** condition (baseline), the backend model receives only the issue description and is asked to resolve it directly, without any explicit planning step.

In the **self-plan** condition, a two-turn interaction is used. In the first turn, the backend model is asked to produce a concise step-by-step repair plan of at most eight steps, without making any code changes. In the second turn, the model receives the plan it generated and is asked to follow it to resolve the issue.

In the **external-plan** condition, the same two-turn structure is used but the planning turn is handled by a separate, stronger model: either a frontier model (Claude Sonnet 4.5 or Claude Opus 4.5, the strongest available Anthropic models at the time the study was conducted and retained for consistency) or a large open-weight model (Qwen3-Coder-480B-A35B-Instruct, hereafter Qwen3-480B). The backend model receives the externally generated plan and is asked to follow it.

This design ensures that the only difference between conditions is the source and quality of the plan. The backend model, the evaluation benchmark, and the scaffold are identical across all three conditions.

## 3.3 Trajectory Collection and SFT Datasets

For RQ2, we collect agent trajectories using GLM-4.7 [39] as the backend model substituted inside Claude Code via DCAS. GLM-4.7 was the strongest available open-weight model at the time of data collection and is a general-purpose open-weight model rather than an SE-specialized one. Task instances are sourced from the SWE-Lego Real Data dataset [24], a curated split of approximately 5,000 resolvable GitHub issues built on top of SWE-rebench [3], which is explicitly decontaminated against SWE-bench Verified so that training on these instances does not leak the test set used in Sections 4 to 6.

Trajectories are collected in a two-phase configuration: the model first generates a plan, then resolves the issue under that plan. We apply rejection sampling to retain only trajectories where the generated patch is verified as correctly resolving the issue, yielding 576 retained trajectories. All Claude Code API interactions are captured using claude-trace [40], an open-source tool that intercepts and logs the full HTTP traffic between Claude Code and the backend model. All collected trajectories are released publicly [26].

Two SFT dataset variants are constructed from the retained trajectories, designed to decompose the contribution of each sense of planning:

- **PlanOnly:** training on the planning phase only (14,162 per-turn samples from 576 instances). Tests whether planning conventions alone transfer without execution context.
- **Plan+Exec:** training on both the planning and execution phases (36,259 per-turn samples from 576 instances). Every trajectory contains an explicit planning turn followed by execution, providing a clean planning-aware signal throughout.

## 3.4 Models and Evaluation Setup

All experiments are evaluated on SWE-bench Verified [6], a human-curated subset of 500 real-world GitHub issues with deterministic solutions, using Pass@1 (the fraction of problems solved correctly on a single attempt). Table 3 summarizes the roles of all models involved.

The execution phase runs for a maximum of 100 turns in every condition, so the budget available for producing and applying a repair is identical across no-plan, self-plan, and external-plan. For self-plan and external-plan, a planning phase runs for a maximum of 500 turns before execution begins. This higher limit is needed because plan generation under Claude Code's multi-turn format can require extensive codebase exploration before a plan is produced. The planning phase makes no code changes, so these turns supply exploration rather than additional repair attempts. The primary training and evaluation scaffold is CC 2.0.76. Fine-tuned models are additionally evaluated on CC 2.1.73, OpenCode v0.0.55, and mini-swe-agent v2.2.8 to assess broader generalization (RQ3).

**Table 3: Models used in this study.**

| Model | Role | Notes |
|---|---|---|
| Claude Sonnet 4.5 | Planner (RQ1) | Frontier plan injection |
| Claude Opus 4.5 | Planner (RQ1) | Alt. frontier planner |
| Qwen3-Coder-480B-A35B-Instruct | Planner (RQ1) | Open-weight planner |
| GLM-4.7 | Source model (RQ2) | Trajectory collection |
| Qwen3-Coder-30B-A3B-Instruct | Executor / SFT target | Primary model |

### 3.5 SFT Training Setup

We conduct full-parameter supervised fine-tuning (SFT) of Qwen3-Coder-30B-A3B-Instruct using LLaMA-Factory [42] with a learning rate of 4.0e-6, cosine scheduler, 65,536-token context, and BF16 precision with Flash Attention 2 and DeepSpeed ZeRO Stage 3. Since Qwen3-Coder-30B-A3B-Instruct is a non-reasoning model, thinking tokens are disabled throughout both training and evaluation. The fine-tuned model weights for both variants are released publicly [25].

## 4 RQ1: How Much Does Plan Quality Matter?

Our hypothesis is that planning conventions, in both the explicit and implicit senses, are a load-bearing scaffold-specific behavior, and that internalizing them through training would close the cross-scaffold gap. Before committing to the expensive intervention of collecting and training on planning-separated trajectories, we first ask a cheaper, training-free question: does planning quality have the kind of leverage that would justify making it the target of training? RQ1 manipulates the most isolatable component of planning, the explicit plan supplied to the executor, under fixed executor model and scaffold, and measures the resulting performance change. If the leverage is small, the hypothesis is unlikely to be worth pursuing. If the leverage is large, planning warrants the methodological investment of RQ2 and RQ3.

### 4.1 Setup

We evaluate Qwen3-Coder-30B-A3B-Instruct under the three planning configurations described in Section 3, using CC 2.0.76 as the scaffold throughout. The backend model, the benchmark, and the turn limit are identical across all conditions, ensuring that any performance difference is attributable solely to the plan.

**Table 4: RQ1 results. Backend model is Qwen3-Coder-30B-A3B-Instruct in all rows. Scaffold is CC 2.0.76. Pass@1 on SWE-bench Verified, max 100 turns.**

| Planning Configuration | Pass@1 (%) |
|---|---|
| No plan (baseline) | 42.8 |
| Self-plan (backend model self-generates plan) | 48.2 |
| Qwen3-480B planner (open-weight) | 49.2 |
| Claude Opus 4.5 planner, fixed selection | 54.2 |
| Claude Opus 4.5 planner, default selection | 56.0 |
| Claude Sonnet 4.5 planner, fixed selection | **57.8** |
| Claude Sonnet 4.5 planner, default selection | **57.8** |

For the external-plan condition, we tested two model selection configurations. In the *default selection* variant, Claude Code's `–model` flag is not set, allowing the scaffold to select models for different internal calls according to its defaults; Claude Code may internally route faster subtasks such as file retrieval to lighter models like Claude Haiku, which could confound attribution of performance to the specified planner. In the *fixed selection* variant, the `–model` flag is set to the specified planner model, forcing all internal calls to use the same model throughout the session and eliminating this source of variation. We additionally evaluated Qwen3-480B as an open-weight planner to assess whether plan quality scales with model capability independently of whether the planner is proprietary.

### 4.2 Results and Analysis

Table 4 presents the Pass@1 results across all planning configurations.

Without any planning step, Qwen3-Coder-30B-A3B-Instruct achieved 42.8% Pass@1 under Claude Code. This is the baseline that reflects the model's behavior when deployed under a scaffold whose planning conventions it has not learned, directly mirroring the cross-scaffold deployment gap shown in Table 1 (Section 2.2).

Introducing a self-planning step raised performance to 48.2%, a gain of 5.4%. This confirms that even without a stronger external planner, structured problem decomposition prior to execution carries meaningful value. The model can partially compensate for the planning gap through its own reasoning.

Providing a plan from Qwen3-Coder-480B-A35B-Instruct yielded 49.2% (+6.4% over baseline), demonstrating that plan quality scales with planner capability even before reaching proprietary frontier models.

Performance increased substantially when a frontier model provided the plan. Claude Sonnet 4.5 achieved 57.8% under both routing variants, and Claude Opus 4.5 achieved 56.0% under default selection, representing a gain of up to 15% over the no-plan baseline. This improvement is attributable solely to plan quality, as the backend model, scaffold version, and benchmark were identical across all conditions.

Regarding model selection, Sonnet 4.5 is unaffected by the `–model` flag (57.8% in both variants), while Opus 4.5 performs slightly better under default routing (56.0% vs. 54.2%).

Claude Sonnet 4.5 outperformed Claude Opus 4.5 as a planner (57.8% versus 54.2–56.0%), despite Opus being the more capable general-purpose model. A potential explanation is that Sonnet's plans are better calibrated to the capability profile of Qwen3-Coder-30B-A3B-Instruct, or that they interact more naturally with Claude Code's scaffold conventions. This finding is discussed further in Section 7.

The 15% planning gain directly contextualizes the cross-scaffold deployment gap in Table 1 (Section 2.2). The SWE-Lego-Qwen3-32B drop of 8.4% from OpenHands to Claude Code is smaller in magnitude than the available leverage from planning. A model trained under OpenHands has not learned Claude Code's planning conventions, and the resulting planning deficit can plausibly account for a substantial fraction of that drop, although the present experiment isolates only the explicit-planning contribution. RQ2 asks whether the implicit planning conventions can also be learned through fine-tuning, and whether the two senses come apart in the data.

**Findings.** Explicit planning is a separable, high-leverage performance component. Providing Qwen3-Coder-30B-A3B-Instruct with a plan from Claude Sonnet 4.5 yields **+15%** Pass@1 on SWE-bench Verified with no change to the executor model or scaffold. Plan quality scales with planner capability, from self-plan (+5.4%) through open-weight planner (+6.4%) to frontier model (+15.0%).

**Implications.** The magnitude of this leverage exceeds the cross-scaffold drops observed in Section 2.2, motivating the methodological investment of RQ2 and RQ3.

# 5 RQ2: Can Planning Be Internalized Through Fine-Tuning?

RQ1 confirmed that high-quality explicit planning helps the model perform well under scaffolds it was not trained for. The natural next question is how to make this planning capability part of the model itself, rather than something supplied externally at inference time. We address this by fine-tuning the model on planning-aware trajectories collected via DCAS, with two dataset variants designed to decompose which sense of planning each composition installs. If training installs implicit planning, the model should improve in the no-plan condition because the conventions are now baked into its turn-by-turn behavior. If training also installs explicit planning, the model should additionally benefit from running an explicit planning turn at inference time. The comparison across variants tells us which sense each composition is teaching.

## 5.1 Setup

We fine-tuned Qwen3-Coder-30B-A3B-Instruct on the two dataset variants described in Section 3: PlanOnly (14,162 per-turn samples from 576 retained planning-phase trajectories) and Plan+Exec (36,259 per-turn samples from 576 retained two-phase trajectories). GLM-4.7 was the source model for all trajectory collection. Claude Sonnet 4.5 and Claude Opus 4.5 were used exclusively as planners for RQ1 and did not contribute to the training data. Full training details are in Section 3.

**Table 5: RQ2 results. Base model is Qwen3-Coder-30B-A3B-Instruct. Scaffold is CC 2.0.76 throughout. Pass@1 on SWE-bench Verified, max 100 turns. Gains in parentheses are relative to the base model without a plan.**

| Fine-tuned Variant | Planning | Pass@1 (%) |
|---|---|---|
| Base (no SFT) | No plan | 42.8 |
| PlanOnly | No plan | 53.8 (+11.0) |
| | Self-plan | 53.2 (+10.4) |
| Plan+Exec | No plan | 52.8 (+10.0) |
| | Self-plan | **55.8** (+13.0) |

## 5.2 Results and Analysis

Table 5 presents the results across the two fine-tuned variants under CC 2.0.76, evaluated under both no-plan and self-plan conditions.

#### 5.2.1 *PlanOnly: training installs the implicit planning conventions.*

The PlanOnly model achieved 53.8% without a plan (+11.0% over base), even slightly exceeding the Plan+Exec no-plan score of 52.8%. A no-plan gain this large from training data consisting *only* of planning turns is direct evidence that what the model has acquired is not the ability to use an explicit planning step (it never gets to use one in this condition) but rather the scaffold's broader structural conventions: the way work is decomposed, the granularity of sub-steps, how tool calls are sequenced, when exploration gives way to action. These are the implicit planning conventions, and they are now active turn-by-turn even without an explicit planning step at inference time.

The same model does *not* additionally benefit from a self-generated planning turn, slightly underperforming with self-plan (53.2%) compared to no-plan (53.8%). This is the predicted converse: PlanOnly's training data taught the model to *produce* plans but never showed it what to do with one during execution, so the two-turn workflow at inference time is unfamiliar and at best neutral. The model has absorbed implicit planning but not the productive use of explicit planning. The two senses are therefore separable in the data.

#### 5.2.2 *Plan+Exec: training installs both senses, with explicit planning adding on top of implicit.*

The Plan+Exec model produced the strongest and most consistent results. Without a plan it reached 52.8% (+10.0% over base), comparable to PlanOnly's no-plan score and confirming the implicit planning conventions have been installed. With self-generated plans it reached 55.8% (+13.0%), a further +3.0% on top of the no-plan score. This brings the internalized result close to RQ1's plan-injection gains: Plan+Exec self-plan (55.8%) matches Claude Opus 4.5 default-routing plan injection (56.0%) and falls 2 points short of Claude Sonnet 4.5 plan injection (57.8%), without requiring an external planner at inference time. The cleanest reading of this differential is that Plan+Exec's training data contained the full two-phase trajectory (explicit plan followed by execution against it), so the model additionally learned the productive use of an explicit planning turn on top of the implicit conventions it already shares with PlanOnly.

The contrast with PlanOnly is the empirical decomposition of the two senses: both variants produce similar no-plan gains because both deliver the implicit planning conventions to the model, but only Plan+Exec adds the explicit-planning benefit because only Plan+Exec contains training data showing how to use an explicit plan during execution.

**Findings.** Planning structure is learnable, and the two training variants cleanly decompose the two senses. PlanOnly installs the implicit planning conventions but not the productive use of an explicit planning turn: no-plan rises to 53.8% while self-plan does not benefit further. Plan+Exec installs both: no-plan reaches 52.8% (implicit) and self-plan adds +3.0% on top to **55.8%** (explicit), matching or approaching the performance obtained with an external frontier planner in RQ1, without requiring an external planner at inference time.

**Implications.** The two senses of planning come apart cleanly in the training data, and both can be internalized through fine-tuning on DCAS-collected trajectories. RQ3 tests whether the combined capability acquired by Plan+Exec generalizes to scaffolds the model was never trained on.

# 6 RQ3: Does the Learned Planning Generalize Beyond the Training Environment?

RQ2 showed that fine-tuning on planning-aware trajectories collected under a single scaffold installs both senses of planning: the implicit planning conventions and, when the training data includes execution against explicit plans, the explicit-planning capability for productively using a planning turn. RQ3 tests whether what was learned is a structural skill that transfers, or scaffold-specific memorization that does not. We evaluate the Plan+Exec fine-tuned model on three deployment targets: a newer release of Claude Code (version transfer), and two open scaffolds the model was never trained on (cross-scaffold transfer). A failure of either dimension would weaken the central hypothesis: if gains evaporate under a newer version, what was learned was release-specific; if they evaporate across scaffolds, what was learned was scaffold-specific memorization.

## 6.1 Setup

Each experiment in this section deploys the same Plan+Exec fine-tuned model from RQ2 into a different evaluation environment and measures Pass@1 on SWE-bench Verified. Training, benchmark, and turn limits are held constant; only the deployment environment varies. The two evaluations below test the two transfer dimensions independently.

**Version generalization.** We evaluate the Plan+Exec fine-tuned model under CC 2.1.73, released after the training data was collected under CC 2.0.76. Both no-plan and self-plan conditions are evaluated.

**Cross-scaffold generalization.** We evaluate Qwen3-Coder-30B-A3B-Instruct (base) and the Plan+Exec fine-tuned variant under two scaffold environments not used during training, via the DCAS interception layer: OpenCode v0.0.55 [22] and mini-swe-agent v2.2.8 [12]. Codex CLI v0.118.0 was attempted but excluded. Codex uses the OpenAI Responses API with SSE streaming rather than Chat Completions, and although our bidirectional translation proxy returned HTTP 200 on all calls, the Codex binary became permanently unresponsive after 10 to 18 round trips and produced no patches.

**Table 6: RQ3 results. Pass@1 on SWE-bench Verified. Base model is Qwen3-Coder-30B-A3B-Instruct. SFT is the Plan+Exec fine-tuned variant. CC 2.0.76 rows reproduced from RQ2 for reference. Gains relative to base model under the same scaffold without a plan.**

| Scaffold | Planning | Base (%) | SFT (%) |
|---|---|---|---|
| CC 2.0.76 (training) | No plan | 42.8 | 52.8 (+10.0) |
| | Self-plan | 48.2 (+5.4) | **55.8** (+13.0) |
| CC 2.1.73 (newer) | No plan | — | 54.6 |
| | Self-plan | — | **57.2** |
| OpenCode | No plan | 46.0 | 47.6 (+1.6) |
| | Self-plan | 47.4 (+1.4) | **49.4** (+3.4) |
| mini-swe-agent | No plan | 35.4 | 39.2 (+3.8) |
| | Self-plan | 37.2 (+1.8) | **42.4** (+7.0) |

All runs follow the evaluation setup in Section 3. The Plan+Exec fine-tuned model was trained only on CC 2.0.76 trajectories and has not seen any of the evaluation environments above during training.

## 6.2 Results and Analysis

Table 6 presents DCAS cross-scaffold results, with CC 2.0.76 rows from RQ2 reproduced for reference.

*6.2.1 Version generalization: gains hold under a newer scaffold release.* Evaluating the Plan+Exec model on CC 2.1.73 yielded 54.6% without a plan and 57.2% with self-generated plans, the highest result in this paper that does not require an external planner. Both scores exceed their counterparts on the training version CC 2.0.76 (52.8% and 55.8% respectively), confirming that the planning behavior transferred robustly to a newer release. This is notable because CC 2.1.73 introduces updated tool schemas and turn management compared to the training version, yet the model adapts without any additional fine-tuning. What was learned is therefore not the surface details of CC 2.0.76 but the underlying planning structure that both versions share.

*6.2.2 Cross-scaffold generalization: gains transfer to scaffolds the model never saw.* On the base model, self-planning already improves performance on all three scaffolds: +5.4% on Claude Code, +1.4% on OpenCode, and +1.8% on mini-swe-agent. DCAS fine-tuning amplifies these gains: the gap between no-plan and self-plan widens after fine-tuning, confirming that planning is the primary acquired capability. Under OpenCode, the SFT model reaches 49.4% with self-plan (+3.4% over base without a plan) and 47.6% without a plan (+1.6%). Under mini-swe-agent, the SFT model reaches 42.4% with self-plan (+7.0% over base without a plan) and 39.2% without a plan (+3.8%).

Across all DCAS-evaluated environments, the SFT model with self-generated plans outperforms the base model without a plan: CC 2.0.76 (training version): 55.8% vs. 42.8% (+13.0%); CC 2.1.73 (newer): 57.2% vs. 42.8% (+14.4%); OpenCode: 49.4% vs. 46.0% (+3.4%); mini-swe-agent: 42.4% vs. 35.4% (+7.0%). The gains are largest under Claude Code, where the structural conventions of the deployment scaffold most closely match what the model internalized during training, and they remain positive on the open scaffolds whose conventions the model has never seen. This is the pattern predicted if Plan+Exec installs a structural planning skill rather than scaffold-specific memorization: the skill carries across, but the magnitude of benefit depends on how closely the target scaffold's conventions align with the structural style the model internalized. A model that had only memorized the training scaffold's surface conventions would degrade on these unseen scaffolds, not gain on them.

**Findings.** The planning capability acquired by Plan+Exec is a structural skill that transfers, not scaffold-specific memorization. It reaches **57.2%** on a newer release of the training scaffold, the highest result in this paper without an external planner, and gains consistently on OpenCode (+3.4%) and mini-swe-agent (+7.0%) under self-plan.

**Implications.** Together with RQ1 and RQ2, this completes the test of the central hypothesis. Planning structure is a load-bearing scaffold-specific behavior that is learnable through fine-tuning on DCAS-collected trajectories, and what is learned generalizes across deployment scaffolds.

# 7 Discussion

The three research questions converge on the central hypothesis stated in Section 2.3: planning structure, in both the explicit and implicit senses, is a load-bearing scaffold-specific behavior behind the cross-scaffold deployment gap, and it can be moved from a fixed scaffold artifact to a learned model capability. This section discusses what these findings imply beyond the particular results, focusing on evaluation methodology, planner-executor calibration, the economics of trajectory collection, and the scaffold-dependence of self-planning.

## 7.1 Single-Scaffold Evaluation is a Methodological Confound

Most leaderboard scores in CLI agent papers are reported under a single scaffold, typically the one the model was trained on. The results in Table 1 show this practice systematically inflates apparent capability: every fine-tuned model in our study degrades on at least one non-training scaffold, and the within-row spread is much larger than the within-row spread of the corresponding base models. The cross-scaffold drops are not noise; they are a real failure mode the field has not been measuring. A claim of the form "model $X$ achieves $Y$% on SWE-bench Verified" is methodologically incomplete without specifying the scaffold and the conventions under which $Y$ was measured.

The natural recommendation is to treat multi-scaffold evaluation as a default rather than an extra. At minimum, a published score should report performance under at least one scaffold outside the training distribution, and ideally as a vector across several scaffolds rather than a single number. Industrial labs have already begun moving in this direction: Qwen3-Coder-Next [4] reports scores under several scaffolds rather than a single number, finding the same limited cross-scaffold transfer our results reveal. As agentic-coding evaluations become more central to the open-model ecosystem, single-scaffold reporting risks consolidating optimization pressure on a single training environment in a way that obscures real generalization deficits.

## 7.2 Planner-Executor Calibration Matters More Than Planner Capability

RQ1 produced a counterintuitive finding: Claude Sonnet 4.5 consistently outperformed Claude Opus 4.5 as a planner (57.8% versus 54.2–56.0%), despite Opus being the more capable general-purpose model. We propose two non-mutually-exclusive explanations. The first is *plan-executor calibration*: plans from a more capable model may implicitly assume a level of reasoning that a 30B executor cannot reliably perform, whereas Sonnet's plans may be more explicit and stepwise. The second is *scaffold alignment*: Claude Code is primarily deployed with Sonnet-class models, and its prompt structure and turn management may interact more naturally with Sonnet's output style. The practical implication is that raw planner capability is not the sole criterion: a planner well-calibrated to the executor and the target scaffold can outperform a stronger but less-calibrated alternative. Future work should study planner-executor calibration as a first-class design variable, including whether the optimal planner changes as the executor model scales.

## 7.3 The Economics of DCAS-Style Training: When Small Data Pays Off

Trajectory collection is the dominant cost in this training paradigm. Each retained trajectory requires a successful end-to-end run of the source model on a task instance, and rejection sampling discards the failures. The economics of training therefore turn on how much trajectory collection is required to produce a useful model. The prevailing assumption in the field is that more is better: SWE-Gym [19], Nebius [28], and CoderForge [2] are at the scale of thousands to tens of thousands of trajectories, all collected under OpenHands. DCAS training uses 576 retained trajectories, an order of magnitude or more smaller than these collections, yet it produces consistent gains across multiple non-training scaffolds.

The interpretation we favor is that the scaffold-portable signal is carried by the planning structure of the trajectory, not the volume of the data. A small set of trajectories that capture both the explicit planning step and the implicit planning conventions of a target scaffold contains, per trajectory, more of the structural signal that this paper isolates than a much larger set of single-pass trajectories collected under a less planning-rich scaffold. The economics follow directly: collecting under a scaffold with rich planning structure, even at small scale, can be competitive with much larger collections under scaffolds without that structure. For practitioners with limited budget for trajectory collection, the choice of *collection scaffold* is therefore at least as consequential as the choice of *collection scale*.

## 7.4 Self-Planning is Scaffold-Dependent in the Base Model and Becomes Scaffold-Generalizable After Plan+Exec Training

RQ3 revealed that self-planning is scaffold-dependent in the base model: it helps under all three evaluated scaffolds, but the gain is largest under Claude Code (+5.4%) and smaller under OpenCode (+1.4%) and mini-swe-agent (+1.8%). The mechanism is that effective self-planning in a base model requires the model to have already internalized implicit planning conventions compatible with the target scaffold; without this, an explicit planning prompt issues a request the model has no idea how to satisfy in the target environment's conventions. Plan+Exec fine-tuning resolves this by installing implicit planning that is structurally sufficient for productive planning under *multiple* scaffolds, and an explicit planning capability the model has practiced using; this is why the no-plan-to-self-plan gap widens after fine-tuning even on scaffolds the model never saw during training. Extending DCAS to other frontier CLI agents such as Codex CLI and Gemini CLI would determine whether the structural style learned from one scaffold is universally useful or specific to that scaffold's design. IDE-integrated and GUI-based coding agents, which typically also communicate with a backend model over an API but expose a different interaction surface, are a further direction.

# 8 Related Work

DCAS sits at the intersection of four lines of prior work: CLI coding agent scaffolds, training trajectory collection, planner-executor architectures, and trajectory-based imitation learning for software engineering.

## 8.1 CLI Coding Agents and Scaffold Frameworks

SWE-bench [10] established the standard benchmark for repository-level issue resolution. SWE-agent [35] introduced the agent-computer interface (ACI), showing that a purpose-built interface substantially improves codebase navigation and editing. Agentless [34] replaces the agent loop with a fixed localize-repair-validate pipeline, demonstrating that separating pre-execution structure from patch generation drives quality. mini-swe-agent [12] reaches above 74% Pass@1 with a strong frontier model despite its minimal harness, illustrating that as model capability grows, scaffold complexity matters less. DCAS adds to this picture by showing that for a fixed scaffold and backend model, plan quality is itself a separable and high-leverage design variable.

Qwen3-Coder-Next [4] collects trajectories from multiple scaffolds, including OpenHands and Claude Code, and independently finds that OpenHands-trained models transfer poorly to SWE-agent while the reverse is moderately successful, corroborating the scaffold-specific bias finding in this paper. Kimi K2 [11] reaches 65.8% via large-scale agentic RL; the DCAS Plan+Exec variant reaches 55.8% at 30B scale with SFT alone, providing a simpler and more reproducible path to competitive performance. SWE-Gym [19] demonstrates that the same training data produces different performance profiles depending on whether OpenHands or MoatlessTools [18] is used as the evaluation scaffold, directly corroborating DCAS's finding that scaffold choice substantially affects performance. Frontier CLI agents such as Claude Code, Codex CLI, and Gemini CLI tightly couple proprietary scaffold with proprietary models, making open-weight substitution difficult. While proxy tools exist in practice [40], DCAS is the first to build a systematic research approach on top of this capability, enabling controlled measurement of planning contributions and scaffold-specific trajectory collection at scale.

## 8.2 Training Data for CLI Coding Agents

Single-scaffold collection has been the prevailing pattern for open trajectory datasets, with few exceptions such as SWE-smith [36], which uses SWE-agent as the collection scaffold rather than OpenHands. SWE-Lego [24] refines SFT with error masking and curriculum training under OpenHands. CoderForge [2] releases the largest open trajectory dataset, also under OpenHands, noting plans to explore multiple scaffolds to improve scaffold generalization. Skywork-SWE [41] achieves 38.0% at 32B and Devstral [20] achieves 46.8% at 24B, both exclusively under OpenHands. DCAS extends this line of work by collecting trajectories under a different scaffold and showing the resulting model gains consistently across multiple non-training scaffolds.

## 8.3 Planner-Executor Architectures

Chain-of-thought prompting [32] established that decomposing tasks into intermediate reasoning steps improves performance, a finding extended by reasoning-oriented models such as o1 [17] and DeepSeek-R1 [9] to agent settings. MASAI [29] extends Agentless's localize-repair-validate pipeline to a multi-agent architecture with a dedicated localization sub-agent, raising file recall from 61% to 75%. CodeR [5] independently confirms the multi-agent localization approach, achieving 28.33% on SWE-bench Lite. HULA [23] provides industrial confirmation: at Atlassian, a dedicated AI Planner producing a coding plan before any code is modified achieved 86% file recall on SWE-bench, and across 663 real JIRA issues engineers approved 82% of the plans it generated. DCAS operationalizes plan-executor separation in a CLI agent context, isolating plan quality as a controlled variable (RQ1) and then internalizing the planner role through fine-tuning so that an explicit external planner is no longer required at inference time (RQ2).

## 8.4 Trajectory-based Imitation Learning for Coding Agents

SWE-Gym [19] established that SFT on expert trajectories transfers behavioral capabilities across model scales. Kimi-Dev [37] shows that Agentless-style training transfers planning and localization skills to SWE-agent frameworks. SWE-RL [33] and DeepSWE [14] demonstrate RL as an alternative to SFT for coding agents. DCAS extends this line of work along a new dimension: SFT on planning-aware trajectories collected under one scaffold transfers across scaffold versions and to scaffolds the model never saw during training, suggesting that planning structure is a transferable behavioral skill rather than scaffold-specific memorization.

# 9 Threats to Validity

We discuss the main threats to the validity of our findings and the steps taken to mitigate them.

## 9.1 Internal Validity

Attributing the cross-scaffold gap to planning structure is supported only in part by our design. Within a scaffold, RQ1 manipulates the plan directly while holding the model, benchmark, and scaffold fixed. Across scaffolds, planning co-varies with tool-call schemas and their parsers, turn management, and context-window handling.

The Pass@1 spread across scaffolds is nonetheless smaller for every untrained base model in Table 1 than for any fine-tuned model, indicating that the divergence is installed by training rather than fixed by the scaffold interface, and the execution phase is capped at 100 turns in every condition, which rules out turn budget as a differing factor. Neither of these fully isolates planning, and ablating each remaining dimension while holding planning fixed is left for future work.

Scaffold version is held constant within each RQ, and the Plan+Exec model is explicitly evaluated on a newer release (CC 2.1.73) in RQ3 as a controlled version generalization test, mitigating version-related confounds.

Training data is collected using GLM-4.7 rather than a frontier model, a deliberate choice showing that gains arise from the scaffold's planning conventions rather than from distilling a frontier model's general capability. Both SFT variants are nonetheless compared against the untrained base model, so the measured gains reflect planning structure combined with general exposure to successful in-domain trajectories. Stronger source models, and controls separating the two contributions such as execution-only, token-matched, or degraded-plan training, are left for future work.

Rejection sampling retains only trajectories where the generated patch correctly resolves the issue, biasing the training set toward easier instances. We mitigate this by retaining 576 two-phase trajectories spanning a diverse range of the SWE-Lego Real Data dataset, and by evaluating on the full SWE-bench Verified set.

## 9.2 External Validity

All experiments use SWE-bench Verified; results may not transfer to other code bases or task types. All experiments also use Qwen3-Coder-30B-A3B-Instruct as the backend model, and the +15% gain from plan injection and the +13% gain from Plan+Exec fine-tuning may not transfer directly to models of different scale, architecture, or pre-training distribution.

DCAS training trajectories were collected exclusively under Claude Code, and cross-scaffold evaluation covers OpenCode v0.0.55 and mini-swe-agent v2.2.8. The planning hypothesis and the cross-scaffold transfer results may not generalize to other frontier CLI agents such as Codex CLI or Gemini CLI, or to scaffolds with different action spaces such as SWE-agent or Agentless. Section 7 identifies extending DCAS to additional scaffolds as a direct future work direction.

Two models in Table 1, SERA-32B and Qwen3-32B-Nex-N1, are fine-tuned to a 32,768-token ceiling that conflicts with Claude Code's hard-coded 32,000-token output limit, causing API failures that do not surface under OpenHands. We resolved this via YaRN RoPE scaling. The incompatibility is a property of these specific released artifacts, not of the OpenHands scaffold in general.

## 9.3 Construct Validity

Pass@1 does not capture trajectory efficiency, turn count, or intermediate reasoning quality. We assess plan quality indirectly through task performance: gains under the self-plan condition could arise partly from improved execution behavior rather than improved planning. Directly scoring fine-tuned model plans is a natural direction for future work.

Claude Code is closed-source and may change across versions. We mitigate this by specifying exact versions (CC 2.0.76 and CC 2.1.73) and by releasing the raw claude-trace HTTP logs so that API interactions can be inspected independently [27].

# 10 Conclusion

This paper started from an empirical observation: open CLI agent models trained on OpenHands trajectories degrade substantially when deployed under other scaffolds, while untrained base models do not show this divergence. We hypothesized that planning structure is a load-bearing scaffold-specific behavior, distinguished into two senses: an *explicit planning* step in which the model produces a plan as a first-class artifact before acting, and *implicit planning*, the structural conventions that shape execution throughout the agent loop. The cross-scaffold deployment gap is, on this hypothesis, a planning-convention mismatch that closes only if planning moves from a fixed scaffold artifact to a learned model capability. We introduced DCAS, an interception layer that substitutes any backend model inside any CLI scaffold without modifying the scaffold, enabling cross-scaffold evaluation and planning-aware trajectory collection that no prior tooling supported.

Three findings together test the hypothesis. *First*, plan quality is a separable, high-leverage determinant of task performance: providing Qwen3-Coder-30B-A3B-Instruct with a plan from Claude Sonnet 4.5 improves Pass@1 on SWE-bench Verified from 42.8% to 57.8%, a 15-point swing that comes purely from varying the plan source. *Second*, planning structure is learnable, and the two senses come apart in training data: PlanOnly installs the implicit planning conventions alone (53.8% no-plan, no further self-plan benefit), while Plan+Exec installs both (52.8% no-plan plus +3.0% under self-plan to 55.8%), matching or approaching the performance of an external frontier planner without requiring one at inference time. *Third*, the learned capability is structural rather than scaffold-specific: it improves further to 57.2% under a newer release of the training scaffold, and gains consistently on OpenCode (+3.4%) and mini-swe-agent (+7.0%) under self-plan, despite the model never having seen these scaffolds during training.

Taken together, these findings support the central hypothesis. The cross-scaffold deployment gap is, to a substantial degree, a planning-convention mismatch; planning can be moved from a fixed scaffold artifact to a learned model capability; and what is learned generalizes across deployment scaffolds. Tool-format and context-length issues remain real and largely orthogonal to the learning problem this paper targets, though we do not isolate their contribution. The path to a single open model whose performance survives a change of deployment scaffold goes through training planning as a structural skill, on trajectories collected under scaffolds that visibly expose the planning conventions one wants the model to internalize, a path that DCAS makes practical at scale.

# 11 Disclaimer

Any opinions, findings, conclusions, or recommendations expressed in this material are those of the author(s) and do not reflect the views of Huawei. AI tools were used for copy-editing. All experiments, analysis, writing, and results were performed by the authors, who also thoroughly reviewed the final content. This complies with IEEE and ACM policies on AI use in publications.

## 12 Data Availability Statement

The DCAS source code and reproduction scripts are available in the replication package on Zenodo [27]. The SFT models [25] and the distilled GLM-4.7 trajectory dataset used for fine-tuning [26] are available on HuggingFace.